# Life, The Mind, and Everything

## The Implications of Incompleteness and Algorithmic Information Theory for Evolution, Consciousness, and Epistemology

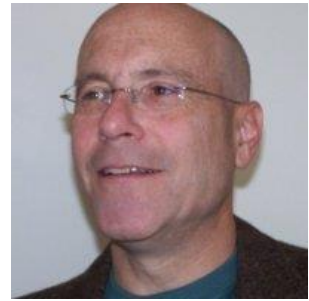

Gary R. Prok

President, Ruby Sneakers, Inc.

Abstract

Incompleteness theorems of Gödel, Turing, Chaitin, and Algorithmic Information Theory have profound epistemological implications.  Incompleteness limits our ability to ever understand every observable phenomenon in the universe.  Incompleteness limits the ability of evolutionary processes from finding optimal solutions.  Incompleteness limits the detectability of machine consciousness.  This is an effort to convey these thoughts and results in a somewhat entertaining manner.

# Life, The Mind, and Everything

"You believe what you want to believe." – Thomas Petty

## Prelude

This is a collection of things I have been pondering for a long time.  I have made every attempt to make this work accurate and unbiased.   I have tried to poke my own holes in my thought processes and correct what could be corrected and discard what could not.  A few of these thoughts are original, but many were learned from others.  The intended reader is one with an engineering or scientific bent and an open mind.  With the intended reader in mind, I allowed myself to use big words without worrying that it would seem like I am trying to impress.  At times it is nice to use big words.

Although some of the concepts presented here are not original, they are stubbornly ignored and remain unassimilated into the philosophies of many scientists and students of science.  It is my hope that this work will present old concepts in a new way to make them more difficult to ignore and will present new concepts to complement the old.

For some time, now I have believed in a principle.  Let's call it The Principle of Divine Undecidability.   It is simply that the existence of God cannot be proven or disproven and any effort to do so is futile.   If God exists, most major religions would be in trouble if The Principle of Devine Undecidability was false.   Religion would unfairly discriminate against those that cannot understand a proof of divine existence.   Only the dull would be damned.  The only out would be another principle where ignorant belief is perfectly fine-tuned to offset inability to understand; anything else would not be fair to the dimwitted. On the other hand, if God does not exist, this Principle means that this non-existence cannot be proven – not even by all the king's horses and men.

Some folks claim Divine Decidability.  Some believe that logic and science can prove the existence of God and all of creation.  Increasingly, others seem to think that science has disproven the existence of God and can explain all that is observed.   I think that these folks may not be quite as Bright as they say that they are.

I hope that reading this will help doubters to see that not everything that is true can be proven.

Some people say the word "faith" like it's a bad thing.  Every worldview requires faith.

## Overture

"Really don't mind if you sit this one out.  My words but a whisper, your deafness a shout." – Gerald Bostock

Since the dawn of The Enlightenment, humankind has climbed a tower of knowledge, clinging to the hope that reason would ultimately offer a way to climb high enough to view and understand all of the wonders of this universe.  Even though great minds of the $20^{th}$ century reasoned that reason itself placed significant and humbling limitations on this quest, we still struggle upward for complete understanding.  Humankind overcomes humbling with hubris and continues to cling to the hope that we will ultimately peer over these limitations and ultimately take our rightful seat as the universal master.

We cannot.

# The Limits of Science

"The truth can't hurt you, it's just like the dark.  It scares you reckless, but in time you see things clear and stark." – Declan McMannus

**Introduction**

Our accumulated understanding of physics, chemistry, biology, mathematics, and all of science, has lifted the veil of mystery from aspects of the human experience that were once thought to be controlled by the supernatural.   We now know that weather is controlled by global wind patterns, regional pressures, the heat of vaporization of water, and many other physical phenomena in a complex ballet.  We know that traits from our ancestors are passed on in DNA.  We know that a thrown ball follows a roughly parabolic path and the variances from that path are controlled by aerodynamics and, to a much lesser extent, general relativity.  Calculus and imaginary numbers allow us to predict the behavior of subatomic particles through quantum mechanics and to calculate the movement of celestial bodies through space-time that is warped as described by general relativity.   By applying this knowledge we can even know the precise location of a person across the world at a specific time from a ubiquitous device carried in a pocket.

The future surely holds more astounding human discoveries.  Many of us look forward to the veil of mystery becoming a rag at our feet once everything that is true is revealed.   Scientific reductionism and materialism are increasingly accepted as the answer to all of life's questions.

But some truths will never be known.

Kurt Gödel proved in 1931 that the limitations imposed by logic prevent us from being able to prove all that is true.  The paradox of the liar, "This statement is false," can be translated into formal mathematics to yield a statement that must be true, but can only be proven so in an inconsistent mathematical system.  A complicated truth of mathematics discovered by Gödel cannot be proved within mathematics, yet humans can prove it to be true.  Although this was disturbing to the mathematical and scientific community, they found themselves able to continue their pursuits of knowledge without much incident.  The only research program that was immediately impacted by Gödel's insight was David Hilbert's attempt to formalize all of mathematics to form a framework for the eventual generation of all mathematical truth.   The goal of this program had to be abandoned and reduced to the forming of a framework for the eventual generation of all that is provable from a given set of axioms.  All that is true would have to be put aside. Yet aside from a few pathological yet true statements that were not provable, the practical impact seemed small.[1]

Things became a bit more serious, though, in 1936 when Alan Turing and Alonzo Church independently built upon Gödel's work and proved that there were mathematical problems that could never be solved.[2]  Turing showed, even before the first digital computer was built, that there was an important problem that a computer would never be able to solve.  This problem was to simply decide if a computer program would eventually give an answer or not.   If one knew that an answer would eventually be forthcoming, one could be patient and wait for it.  Turing proved that waiting was a potentially fruitless option.  There is no way to program a computer to analyze another general computer program, or programmed sequence of logical steps, and determine in advance that program would not just continue, step-by-step, for all eternity.[3]

Things became even more serious in 1974 when Gregory Chaitin proved, using Algorithmic Information Theory, that there are an infinite number of mathematical truths that cannot be proven. The unknowable very quickly went from one confusingly worded statement to an infinite number of statements. The knowable went from a potentially complete understanding of truth to a mere subset of all that is true, complemented by an infinite number of unknowable truths.[4]

**Rice's theorem**

Even though the halting problem and variations to it can be considered an extension to Gödel's work, it is worthwhile to consider them first, since they are conceptually easier to visualize. Turing proved that there is no general way to determine if a particular computer program operating on a particular input will ever halt. Instead of merely considering how to determine halting, it is useful to consider how a one might determine if a computer program might do any specific thing, such as halting or printing a specific output string.

Rice's theorem can be used to extend Turing's theorem to detecting computer behavior beyond halting. A computational corollary to Rice's theorem states that any non-trivial property of a computer program cannot be decided by a computer program.[a] (A non-trivial property of a computer program is a property that not all programs exhibit). The following graphical proof technically applies to programs that compute an output from an input, while Rice's theorem technically applies to functions that produce one numerical string from another numerical string. However, proving Rice's theorem for computer programs is equivalent to proving it for functions operating on numerical strings.[5]

**A graphical explanation of Rice's theorem**

A non-trivial property of a computer program is manifested as output *O* on an input *I*.

Suppose a computer program, **r**, exists that is given as its input any computer program **P** written in a prescribed language, along with any input, *I*, that the program will be given to process. The program **r** will interrogate the program **P** and input *I* and determine whether or not the program **P** given input *I* will output *O* or not. Note that **r** must be a properly coded program; it cannot be a nonsense string. It is assumed to be what is called "well formed."

Now consider an extension to program **r**, named **R**. Program **R** only has one input that it uses both as an input program **P** and as the input *I* to that program. It takes the program **P** input, stores it, and then executes program **P** using the stored version as the input. Program **R** is constructed to print the output *O* and to halt if it detects that input program **P** does not output **O**. Program **P** simply loops if it detects that input program **P** outputs **O**.

---

[a] Note that certain straightforward programs can be proven to halt on any input; Turing's proof is a statement about programs in general. Likewise, certain straightforward programs can be proven to generate a specific output on any input; Rice's theorem as applied to computer programs is a statement about programs in general.

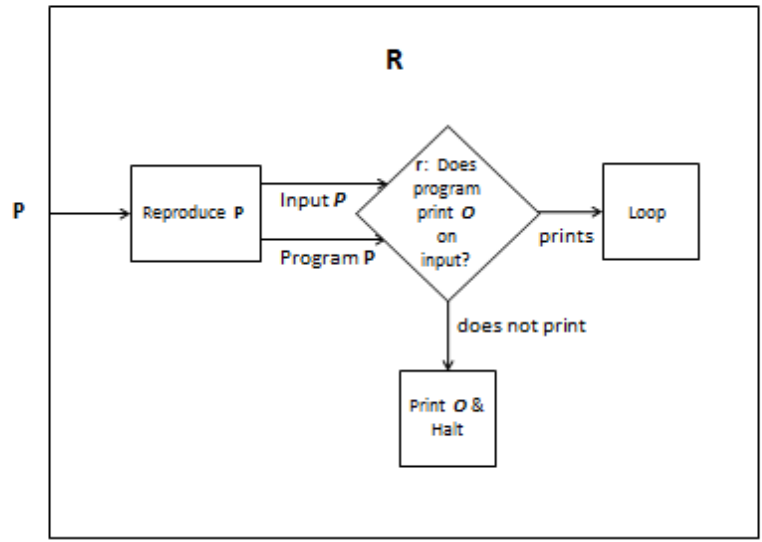

Figure 1 – Construction for proving Rice's Theorem

Program **R** can be given its own code as its input. Since **r** is well formed, **R** is also well formed and acceptable to use as an input. (This is the actual embodiment of what program **R** does – to take a program and use it as its own input).

This interesting construction presents a logical challenge. If program **R**, given its own code as input, prints the output **O** and halts, then program **R**, given its own code as input, does not output **O**. If program **R**, given its own code as input, does not output **O**, then program **R**, given its own code as input, outputs **O**. Both cases are logical contradictions. The only way out of this contradiction is if the supposition that program **r** exists is false.[b]

There does not exist a program, **r**, that can always determine if program, **P**, outputs **O** on input **I**.

**A graphical explanation of Gödel's Theorem**

---

[b] Note also that this implies that even input free programs cannot be interrogated to determine whether or not they output **O**. If they could, then one could replace a program **P** and input **I** with a new program with the input hard-coded as part of the program and in that way determine if **O** is output, which cannot be done.

The Halting Problem and Rice's Theorem give a glimpse of things that are mathematically true and remain unknowable. They may represent eye-opening truths to students of mathematics, engineering, or computer science, but seem to be limited to specific problems in a small area of research. Gödel's Incompleteness Theorems show that this area is actually quite large.

Gödel's First Incompleteness Theorem proves the inability of any sufficiently powerful mathematical system to prove everything about that system that is true. A sufficiently powerful mathematical system is one capable of expressing elementary arithmetic. Here we will use the term sufficiently powerful to mean a mathematical system powerful enough to represent any computer program. Gödel actually proved two incompleteness theorems; the second being a corollary to the first. Here we only consider the first.

Consider a set of axioms that describe the basic rules of mathematics and a set of logical steps that allow one to combine and cascade application of these axioms to create mathematical truths. Together these constitute a formal system. The mathematical truths resulting from such a cascade are theorem statements. A proof is a cascade of axioms and logical steps that is paired with the final theorem statement. The formal system of axioms and logical steps is *complete* if all true mathematical statements can be generated by a set of such cascades and thereby proven. The formal system of axioms and logical steps is *consistent* if no two theorems generated by a set of such cascades contradict each other. Mathematics is generally believed to be consistent. If mathematics is not consistent, then it quickly crumbles into a useless formal system, a system that can prove anything, including 0=1.

It is possible to perform an orderly search of every possible cascade of applied axioms and logical steps to generate a sequence of mathematical truths, or theorems. This orderly generation of theorems will never end, with theorems growing larger over time, with each new step branching off to many possible next steps. The process of enumerating theorems and proofs is called a British Museum Algorithm, since a procedure similar to this was described as a conceptually sound, yet impractical, means of producing all of the books in the British Museum. In 1920, David Hilbert proposed such a process to generate all mathematical truths.

An important concept in Gödel's theorem is the free variable. In mathematics, a theorem can have a free variable, say "x." For example a theorem could be:

> There exists a number x, such that 3 = 1+x

A theorem can also involve a procedure. For example a theorem could be:

> For any integer x, expressed in base-10, x is divisible by 3 if, and only if, the sum of the digits is divisible by 3.

The set of orderly rules for theorem generation allows for such constructions.

This process of theorem discovery can be automated in what is familiar today as a computer. The formal system of axioms and set of logical steps can be encoded into a symbolic system so that any proof can be represented by a string of characters. The orderly search of all valid strings can be programmed into an algorithm to be executed on a computer, and theorem-plus-proof after theorem-plus-proof could be generated. The resulting theorems output by the computer are also uniquely encoded as strings of characters; strings which translate into mathematical words and symbols which we are familiar with.

A proof could translate to contain phrases like:

> Add to y the largest integer less than (x divided by 10 to the power (the largest integer less than the Base-10 logarithm of x)) , or, in another language:

> y = y + Int(x/10^Int(Log10(x)))

This step is used to determine the value of the first digit of an integer and could be part of a proof of the example theorem statement:

> For any integer x, expressed in base-10, x is divisible by 3 if, and only if, the sum of the digits is divisible by 3.

A proof and theorem discovered this way could include statements having loops and conditional if-then statements, like statements of a computer programming language.

The interesting thing is that a numerical representation of a theorem can actually be used as an input to any theorem that has a free variable. Any integer can be substituted for a free variable, including the coded output string representing a theorem. Gödel recognized this ability of mathematics before a computer was invented and invented a way to give every mathematical statement a unique "Gödel" number.

Further, this encoded theorem can be expressed in the same computer language used to encode the search algorithm. If it is, this allows for interesting constructions, one of which is presented later.

It is straightforward that a program could be written to take a numerical representation of a theorem as an input and use the British Museum Algorithm to generate theorems until one matching the input or one matching the negation of the input is discovered. In this way it could be decided if a candidate theorem is provable or not-provable, assuming arithmetic is complete.

It is also straightforward that a program could be written to only decide if candidate input theorems are **not** provable, by searching until only the negation of the input is discovered. A candidate theorem that is provable, in a consistent system, will cause this program to look continuously and fruitlessly for proof that the candidate theorem is not-provable. Call this program **"b"** and the input candidate theorem **"T"**. The input **T** can be any properly formed mathematical statement, including any computer program in a chosen language.

Program **b** can be made more flexible by allowing the input to be theorems having a single free variable. Call this variable **X**. Now inputs **T** and **X** define the theorem that is to be disproven. For example **T** could be "3 = 1 + **X**", and **X** could have the input value of 1. This would then be disproven by **b**. In this case, program **b** will accept the input **X** and the theorem candidate **T** and determine if there is a proof that they do not constitute a valid theorem. It will eventually prove that 3 does not equal x+1, where x=1, if mathematics is consistent.

Now consider an extension to program **b**, named **"B"**. Program **B** only has one input that it uses both as its input theorem and as the free variable value to be used in that theorem. Program **B** prints the proof that Theorem **T**, with free variable *T* is not a valid theorem, if it finds such a proof; or program **B** will search indefinitely if no such proof exists.

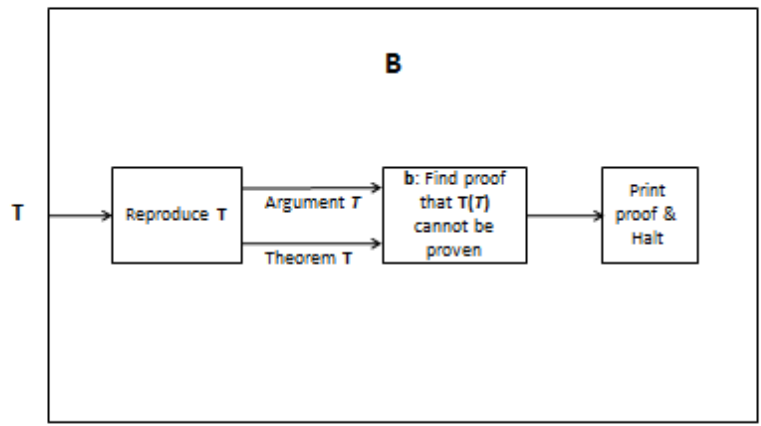

Figure 2 – Construction for proving Gödel's First Incompleteness Theorem

Another construction similar to **B** that is a parallel construction to that used above for the Rice's Theorem:

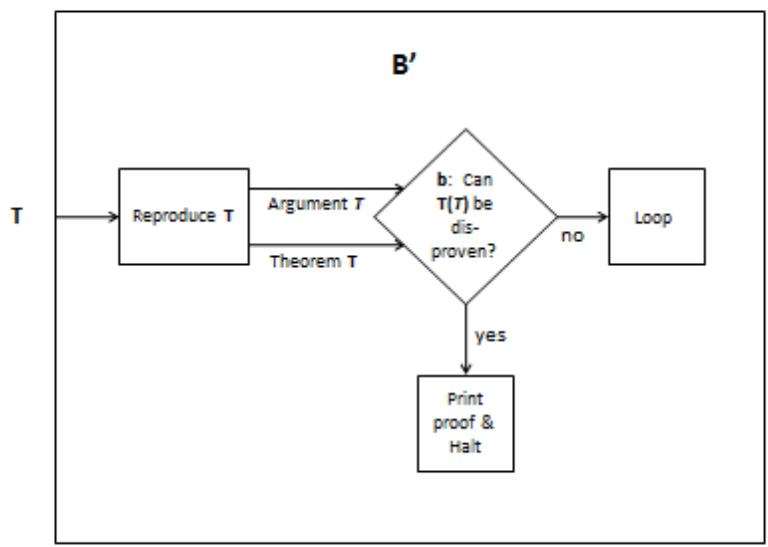

Figure 3 – Parallel construction for proving Gödel's First Incompleteness Theorem

Now consider further the case where candidate theorems are expressed in the same computer language used to encode program **B**. This means that any properly coded computer program can be considered as a candidate theorem. The candidate theorem that is input can be any program including program **B,** and program **B** can be given its own code as input. (This is the actual embodiment of what **B** does – to take a theorem candidate requiring an input string, and use its own representation as its own input). In this case **B** is just an extension of **b**, and **b** is an encoded statement that delineates the provability of any candidate theorem having a specifically defined free variable.

This interesting construction presents a logical challenge; it is a formal version of the Liar's Paradox: "This statement is false.**"** It is here stated interpreted as "This theorem is unprovable". If the computer finds that **B**(*B*) is not a provable theorem, then the formal system is inconsistent; it proved itself as non-provable, which is inconsistent. If the computer never finds that its input is not a theorem, it means that the formal system is not complete; the input language is non-provable, but cannot prove that fact. This means that program **b** does not exist, if mathematics is consistent. Therefore mathematics is either inconsistent or incomplete. This is the conclusion of Gödel's Incompleteness Theorem.

At a higher level, or meta-level, this means that the theorem represented by this interesting construction is true, but unable to be generated, or proven, in a consistent formal system. It is true, but not provable, that **B**(*B*) is not a provable theorem. There exits at least one true but not provably true mathematical statement. This places a significant limitation on mathematics.

This limitation profoundly changed the philosophy of mathematics. Hilberts' program to theoretically mechanize all of mathematics was doomed to fail. It was now known that all that is true could never be known. The veil will always hide some mystery.

One can still attempt to explore for new truths by taking a conjecture that is intuitively suspected to be true and unprovable, and making it a new mathematical axiom. One such conjecture is Goldbach's conjecture, one of the oldest unsolved problems of mathematics:

>  Every integer greater than 2 can be expressed as the sum of two prime numbers.

Adding this as a new axiom to a formal system of mathematics opens the door to new mathematical truths, based on the significant assumption that this new axiom is consistent with the other axioms. It also opens the door to potentially inconsistent mathematics. This introduces a level of experimentalism to mathematics. Chaitin has argued that mathematics needs to borrow a page from the playbook of the sciences and start using the scientific method to advance mathematics. Hypothesizing that the new axiom is true requires faith; faith in its consistency within the formal system.

Since the time of Gödel, other mathematical constructions have been found that are true, but not provable. Gregory Chaitin has extended Gödel's argument using Algorithmic Information Theory to prove that there are infinite mathematical truths that cannot be proven. They merely exist.

**Algorithmic Information Theory**

Algorithmic Information Theory (AIT) is a branch of mathematics that explores the amount of information contained in a numeric string and how information is passed through computer programs or algorithms.  The amount of information in a string increases as its complexity increases.   The binary string "0000000000……", where there are a million zeros, is not very complex.   One could write a short program is a chosen programming language to print "0" and loop back and do it again until it counts that it did this a million times.  This would be a short program that would not require many bits of information.   On the other hand the string "11101001101…..", where a coin is flipped a million times and "1" represents heads and "0" represents tails, would be difficult to compress into a short program that generates it exactly.   The program would have to contain about a million bits of information.

The complexity, formally the Kolomogrov complexity, of a string is the length in bits of the smallest input-free program that generates that string, and only that string, in a chosen programming language.   Simple strings only require short programs; complex strings require long programs.  In a similar way, a digital photograph of a white wall can be compressed into a much smaller file than can a digital photograph of a Jackson Pollack painting.  The white wall would require a smaller input-free program to generate it than would the Jackson Pollack painting.

For extremely large and complex strings the choice of programming language does not make significant difference in the complexity measure.  Since one language can be translated to another using an interpreter having a finite complexity, a minimal program in one language can't be any bigger in another language than the size of the interpreter.   It might even be smaller; an interpreter going the other way bounds how much smaller it could be.  Again it can't be any smaller in another language than the size of the interpreter.  Language choice changes the complexity measure by no more than +/- some finite number.   This is the Invariance Theorem of AIT.[6]

The complexity of a large string cannot be computed.  Its complexity is an unprovable truth, like the unprovable truth that Gödel discovered.   Given an encoded algorithmic system, suppose an input-free program that uses that system can be written with a finite complexity, F, which generates proofs in British Museum fashion until it finds a proof that a particular string, N, has complexity greater than F.  Then the program prints "N" and halts.  But this is impossible; the program would generate a number with complexity higher than that of the program that generated, which violates the definition of complexity.  The only way out of this contradiction is if there is no proof that an integer has complexity greater than F.  This is Chaitin's Incompleteness Theorem.[7]

The complexity of a sufficiently complex string is incomputable and therefore unknowable within the formal system of mathematics on which programming is based.  There are an infinite number of strings with complexities that are unknowable.

Even some strings of complexity less than or equal to F cannot be determined.  This is illustrative even if limited to finite strings.  If all finite strings having complexity less than or equal to F could be determined, then all finite strings having complexity greater than F could be determined by elimination.  But there can be no proof that a particular string, N, has complexity greater than F. Therefore even some strings with complexity less than F cannot be determined.

The value of F depends upon the language chosen, as does complexity in general. Computer languages also depend upon the computer processor architecture used. All generally useful computers are Turing Complete, meaning that they can be simulated on a Turing Machine, an extremely simple, but time-inefficient computer concept. So a meaningful upper bound of F would be the complexity of a computer program as described above which runs on a Turing machine. This estimate would be the complexity of the computer program above in a given language, plus the complexity of its language's complier to run that language, on a simple Turing machine. An estimate of the total complexity of these is less than 100 kilobytes, which is an upper bound of the complexity of F. No string with complexity greater than 100 kilobytes can be proven to have complexity greater than 100 kilobytes, and not all strings with complexity less than 100 kilobytes can be proven to have complexity less than 100 kilobytes.[c]

There are countless mathematical truths that cannot be proven or known.

## Interlude

"A man hears what he wants to hear and disregards the rest." – Paul Simon

These proofs may seem academic, arcane and perhaps insignificant. However, many academics promote a view of reductionism. This is the notion that everything and every phenomenon in the universe can be explained by digging deeper and deeper into the underlying mechanisms until the simplest and most fundamental underlying mechanisms are uncovered. Once these fundamental mechanisms are understood, then the rest of the explanations follow. Animal behavior is a biochemical mechanism, biochemistry is explained by physical chemistry, physical chemistry is explained by quantum physics. Quantum physics explains all of the above. The only things that determine animal behavior that cannot be explained are fundamental constants of physics, which are being measured ever more precisely.

---

[c] The British Museum program above has an estimated program size of less than 10 kilobytes in a language like language BASIC. A small version of BASIC requires only about 4 kilobytes to run on a typical desktop computer. A program whose abbreviated name is BF is a small Turing machine language requiring less than 1 kilobyte, and a BASIC to BF complier is less than 80 kilobytes long. These sum to 95 kilobytes and their complexities sum to a smaller number. So it is safe to say that the complexity limit, F, is less than 100 kilobytes, or 800,000 bits.

The notion that science will eventually explain every observable phenomenon is becoming pervasive. Thoughtful lay people tend to accept these views as coming from experts and are unable to find a reason to doubt them.

There is reason.

Physics, including quantum physics, is understood through mathematics.  Mathematics is incomplete, it cannot even tell us if a relatively simple computer program will give an answer or not.  Our understanding of physics could be embellished with new results and new explained phenomena, if only we knew what new mathematical axioms to add.  Yet, even though there are countless unprovable mathematical truths, these axioms can only be surmised.  They cannot be derived.

Similarly advances in physics are built on conjectures.  These are reasonable assumptions that are not proven from first principles.  These conjectures may actually have parallel conjectures in mathematics.  An example is the Odlyzko-Montgomery Law which states that the statistics of energy levels of a large atomic nucleus are the same as those of the Riemann zeta function.  The Riemann zeta function in turn has statistics that are thought to follow the Riemann Hypothesis, the truth of which is today undecided.[8]

Theoretical physics employs esoteric mathematics, many of which were developed by mathematicians long before any practical application was found.  It would require a tremendous leap of faith to assume that none of the unprovable truths of mathematics have any bearing on the predictive power of theoretical physics.[9]

The belief that a mathematical description of physics can be complete and can explain everything is just that, a belief that is faith-based.

# The Limits of Evolution

"The learning and knowledge that we have, is, at the most, but little compared with that of which we are ignorant." – Plato

**Introduction**

The diversity of life and the extraordinary adaptations to unique and even extreme environments is seen as a testament to the success of mutation and natural selection as a way to optimize any thing for any purpose. Evolutionary concepts have been used to design devices that work very well even though the exact mechanism of how they work is non-intuitive and puzzling. There is a tendency to believe that the best solution to any design problem can be generated by a strong enough evolutionary technique. A corollary is that every possible life form can be generated by evolution.

**Genetic Algorithms**

An example of how the limits of knowledge impact our practical use of science is related to genetic algorithms. Genetic algorithms are a type of optimization technique. Optimization is the process of searching for a best alternative among many. Optimization is used to schedule airline flights, pick a most efficient route to travel through every desired city, choosing which house to buy, and choosing where to invest one's money, as some examples. Optimization techniques can be as poor as flipping a coin, or, for some specific problems, as good as always being able to find the absolute best alternative.

A genetic algorithm can be understood through its embodiment as a computer program. In simple terms it is a program **GA** that takes an input string **E**, and outputs the value of an internally manipulated string **Gn** which produces a maximum value for a calculated variable **Obj**. **Obj** is calculated, given **E** and **Gn**.

> Program **GA** manipulates the string **Gn** from one trial to the next in a manner that mimics or is inspired by evolutionary changes in the genetic code of a life-form from generation to generation. The string **Gn** is analogous to a genetic code, or genome. String **Obj** is the value of what is called an objective function that is to be maximized. Typically **Obj** is a measure of the fitness of a calculated phenome, **Ph**, or physical object, resulting from the genome. **Ph** is calculated by program **P** using input **Gn**. Input **E** is a string defining the environment in which the phenome is simulated to exist. For example, the phenome may be a simulated radio antenna whose shape, encoded in **Ph**, is determined by string **Gn**. And the environment **E** may encode a radio wave to be detected by an antenna phenome. The fitness **Obj**, of an antenna would be the magnitude of the calculated voltage induced in the antenna by the radio wave defined by string **E**.

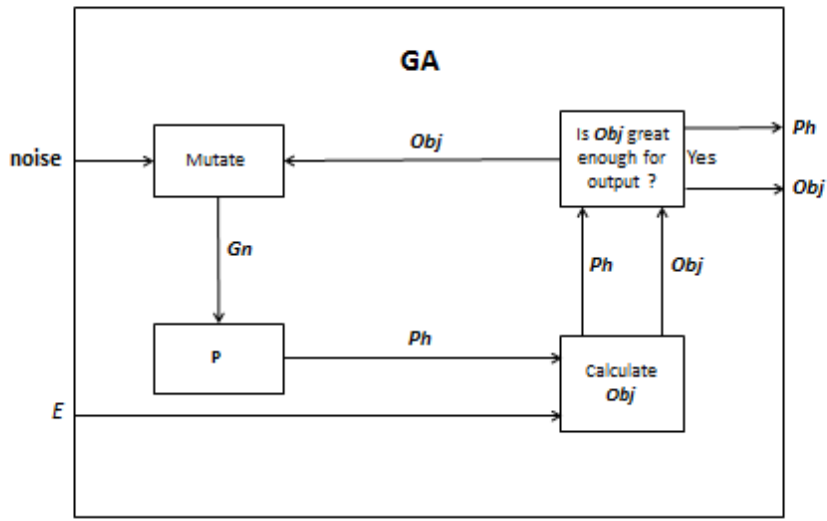

Figure 4 – Construction for proving the limits of an evolutionary process

The genetic algorithm **GA** has rules for changing the string **Gn**, based on the values of **Obj** for previous trials of **Gn**. The most-fit trials of **Gn** are selected and used to generate new trials of **Gn**. A change in **Gn** could be a simple one-bit change, or mutation, or it could be more complex and include blocks of code being inserted from one **Gn** try into another. Mutation could lengthen or shorten the string as well. The type of mutation and the details of the mutation are determined randomly. If too many random mutations are allowed at once, the genetic algorithm approaches a random search algorithm, and loses its advantage, so only a few mutations are allowed for each trial of **Gn**. Although pseudo-random number generators are generally used, a truly random mutation will require a second input of random bits, or noise, to drive the mutation process.

Typically, entire generations of strings **Gn** are generated and evaluated before a new generation of trial strings of **Gn** are generated for evaluation. (**Gn**, **Ph** and **Obj** would represent many instances in the diagram above). Also, typically any trial string **Gn** meeting a minimum threshold of fitness is output by **GA**.

As with programs in general, Rice's theorem places a limitation on genetic algorithms.

**Rice's Theorem and Genetic Algorithms**

Taken as a black box, program **GA** takes one input string **E** and generates a series of output strings **Ph**$_i$. Rice's theorem suggests that it is impossible to know if **GA** will output a particular instance of **Ph**. (There does not exist a program, **R**, that can determine if any program, **P**, outputs **O** on any input **I**). By analogy, one might say that there does not exist a program, **R**, that can determine if any program, **GA**, outputs a particular desired instance **Ph** on any input **E**. This implies that a genetic algorithm cannot be analyzed and proven to guarantee to output a specifically expected good solution. The validity of this implication depends upon the nature of program **P** and how it calculates phenome **Ph** .

Many genetic algorithms have a straightforward way to map a genome to a phenome, where Rice's Theorem would not apply. The first byte of a string **Gn** could specify the length of the first segment of the antenna phenome. The second three bytes could specify the angle in three dimensional space of the next segment relative to the first. The next byte could specify the length of the next segment, and so on. This section of the genetic algorithm is guaranteed to complete and halt, and allow the algorithm to proceed in a short amount of time to the next step. These straightforward genetic algorithms are not susceptible to lengthy, looping calculations while generating a phenome. This also prevents them from finding some creative solutions. Although the antenna example described here limits the configuration to that of an intricately bent coat hanger, a more general genetic algorithm could allow for an infinite variety of configurations, like ferns, cages, snowflakes and any shape imaginable. Rice's theorem does not apply to programs as simple as this less general algorithm.

A more general genetic algorithm would use a more complex calculation to generate a phenome from a genome. Block **P** would entail calculations involving loops and conditional branching statements; statements containing "go to" and "if-then-else." In nature, the translation of the genome to the phenome is a complex process that includes conditional branching; parts of the genome may be skipped dependent upon the previous read DNA and on chemical signals generated from the partially assembled phenome. Such a genetic applies to such a genetic algorithm. Such a genetic algorithm is not guaranteed to output a specific phenome that is known to have a high valued objective function.[10][11]

Yet one might be able to engineer **GA** to ultimately produce a particular instance of **Ph** that is known to be successful.[d]

So, it is conceivable that a sub-set of genetic algorithms, with very particularly chosen input strings of random noise digits, do indeed output a particular instance of **Ph,** given enough time. Such algorithms amount to teleological, or directed, evolution.

Alternatively, with enough random mutations introduced to some trials of **Gn**$_i$, the entire space of possible **Gn** strings could eventually be explored and a desired version of **Ph** might be guaranteed to eventually be generated. A low-bar acceptance threshold of acceptable fitness would allow many mutation branches to be explored. This would amount to changing the genetic algorithm process into that of a random search and the value of using a genetic algorithm would be lost. If program **P** actually halts and gives a result for every input, **Gn**$_i$, then Rice's theorem would no longer apply; every possibility could eventually be tried. A problem for random searches is that, even for modest sized search spaces, the number of possible solutions is exceedingly large; a 500 bit genome would have $10^{150}$ possibilities. It is impractical for the fastest theoretical computer to search them all in a time span equal to the current age of the universe.  Such algorithms are not practical or useful and are considered computationally intractable.

---

[d] If one holds the environment string **E** fixed, and allows the input string of random bits to vary from one long, but finite, run of the genetic algorithm to sufficiently many subsequent long, but finite runs, then eventually a particular desired instance of **Ph** might be output.

The limitations to genetic algorithms described here are significant.  Yet it is comforting to think that a realistic genetic algorithm has a reasonable chance of discovering any particular imaginable unique genome and any particular imaginable phenome.  Some of these genomes could produce phenomes with configurations of exceptional fitness.  Any antenna shape, including of cages of ferns covered with snowflakes could be generated.

Algorithmic Information Theory (AIT) shows that this cannot be the case.

**AIT and Genetic Algorithms**

The result of AIT that is germane to the limits of genetic algorithms is that the complexity of a string of sufficiently high complexity cannot be computed, per Chaitin's Incompleteness Theorem.    This fact that high complexity strings have indeterminable complexity has limiting implications for genetic algorithms.

Suppose we have a genetic algorithm, **GA** that has the objective to form a phenome that is a perfect complement of the string *E*.   It will search for an input *Gn* program that generates a phenome string *Ph* that is a complement to *E*, with every "1" of *E* replaced with a "0" -  and every "0" of *E* replaced with a "1".  The complement string will have nearly the same complexity as *E* and this complexity will not exceed the number of bits in *E* by very much.

The complexity of *Ph* is related to the smallest size *Gn* that will produce the complement to *E*.  Specifically, the complexity of *Ph* is equal to the complexity of *Gn* plus the complexity of the simulation program **P** that generates this complement phenome.

Many different *Gn* strings can result in the same phenome.  In nature two such strings could simply differ in their "junk DNA", or they could conceivably generate same phenome by different means.   However, there is a minimal length *Gn* that is determined by the complexity of *Ph*.  Suppose that **GA** can find every possible *Gn* that produces the complement to *E*.  If it could, then it could also find the minimal length *Gn*.  However, it cannot be proven that **GA** generates this minimal string for a sufficiently complex *Ph*.  If it could be proven, then one could use **GA** to compress any string to its minimum and thereby determine its complexity.  This violates the Chaitin's Incompleteness Theorem, which states that high complexity strings have indeterminable complexity.  Therefore **GA** cannot find every *Gn* that produces the complement to *E*.

There exist *Gn* strings that produce a given phenome that cannot be proven to be generated by a genetic algorithm.

**Introduction of New Genetic Information**

This opens the door for specific and successful compressed versions of *Gn* strings to be externally introduced into a genetic algorithm; strings which the genetic algorithm cannot generate on its own. These strings could then be used in subsequent generations to generate entirely novel trials of *Gn* resulting in entirely novel phenomes. It is conceivable that these novel phenomes could have unprecedented success.[e]

In nature, one can imagine stumbling upon valuable genetic information from another organism originating in a foreign environment. This genetic material is then taken up by micro-organisms which then spread it to higher organisms through horizontal gene transfer. The notion here is that this string is not random, but was generated by an evolutionary process in some very foreign environment, where it enjoyed success. Through a *Junkyard Wars* style repurposing, this genetic material might well find new uses in the new organism, in the new environment.

What would truly make adding new genetic material useful would require some intuitive understanding of what *Gn* strings would be useful. This is analogous to the notion that mathematicians should use their intuition to add new axioms to formal systems to allow new mathematical truths to be explored. Unfortunately we do not yet understand existing genomes well enough to be able to add useful genetic information to systems as complex as those in life on earth.

It took knowledge greater than that contained in a formal system of mathematics, i.e. meta-knowledge, to generate Gödel's Incompleteness theorem. It takes similar meta-knowledge to generate intuition required for adding potential axioms to a formal system of mathematics. It would also take similar meta-knowledge to add new genetic information to a genetic algorithm in order to generate phenome configurations that the genetic algorithm would not otherwise be able to explore and improve upon.

**Implications**

Many students of science are materialists and believe these principles as a matter of faith:

- Anything that can be observed is explainable.
- Any explanation of observed phenomena only involves material things.
- All interactions of material things can be quantified mathematically.

---

[e] If one were to somehow stumble upon such a compressed genome yielding a phenome of good fitness, it would be similar to adding a new axiom to a formal system. In a formal system, this can be done by taking a conjecture that is intuitively suspected to be true and making it a new mathematical axiom. Both cases would involve addition of new information to the system or process. As discussed earlier, adding a new axiom to a formal system of mathematics opens the door to new mathematical truths based on the assumption that this new axiom is true. Similarly, adding a new genome to the mix of one generation's trials of *Gn* could open the door to new phenomes of good fitness.

Therefore:

- Any observed complex phenome, be it a plant, animal, fungi, or other eukaryote, or bacteria, or archaea, is explainable.
- This explanation only involves interactions of material things with a material environment.
- The interactions resulting in the final plant, animal, or fungi can be modeled using a complex genetic algorithm on a computer if one had sufficient knowledge.

However, not all possible phenomes can be produced by a genetic algorithm.  So it is a matter of faith to believe that every existing phenome, be it plant, animal, fungi, or other eukaryote, or bacteria, or archaea, could be the result of material things interacting with one another.  It is even a stronger matter of faith to believe that every existing phenome is indeed the result of material things interacting with one another.

This does not mean that every living thing, including the species *homo sapiens*, did not arise through material interactions.   Rather, it requires faith to believe that this is how they all came to be.  Belief in Dawkin's "Blind Watchmaker" requires blind faith.

# Consciousness

"I think, therefore I am."– R. Descartes

**Introduction**

Most people seem to believe that they are conscious. Most of these people believe that other people are conscious as well. Materialists believe that someday we will be able to create machine consciousness.

There are, however, limitations of this worldview.

**Detection of Consciousness**

Consciousness is difficult to define and pin down. Here we simply state that consciousness is the state awareness and of being aware of one's state of awareness. Per this definition, a being that has this awareness would also know that it is conscious.

Suppose that a computer program can be written which displays consciousness. A conscious program would become conscious by virtue of its code, so the code must hold the key to its consciousness. This suggests that the limitation that Rice's theorem imposes on computer programs might apply.

Suppose a non-conscious computer program, **c**, exists that is given as input any computer program **P** written in a prescribed language, along with any input, **I**, that the program **P** will be given to process. The program **c** will interrogate the program **P** and input **I** and determine whether or not the program **P** given input **I** will exhibit consciousness or not. Note that **c** must be a properly coded program; it cannot be a nonsense string. It is assumed to be what is called "well formed."

Here **I** is a finite string. After the string **I** has been processed, no new information would be presented to program **P**, any consciousness might find itself rather bored, as if were in an isolation tank alone with its thoughts. It could still be pondering, however, unless it ends up halting.

Now consider an extension to program **c**, named **C**. Program **C** only has one input that it uses as both its input program **P** and the input **I** to that program. It takes the program **P** input, stores it, and then executes that input program using the stored version as its own input. Program **C** is constructed to exhibit consciousness and then halt if it detects that input program **P** with input **I** does not exhibit consciousness. Program **C** simply loops if it detects that input program **P** with input **I** exhibits consciousness.

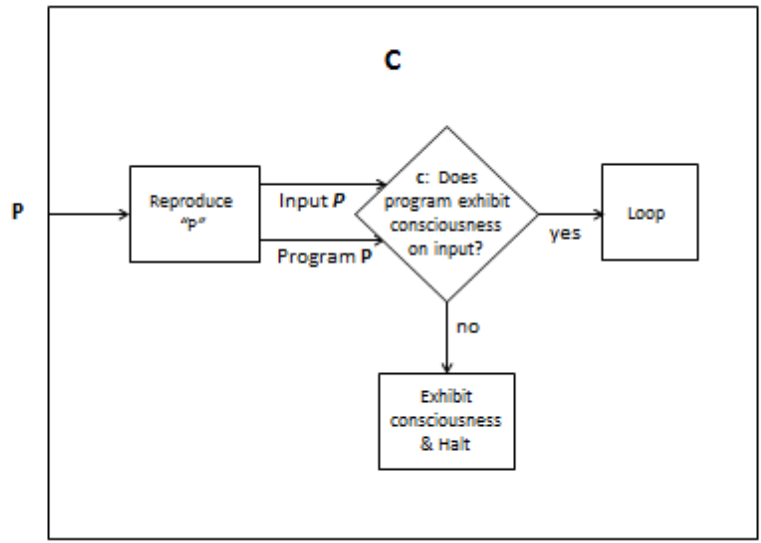

Figure 5 – Construction for proving the limits of machine consciousness

Program **C** can be given its own code as its input. (This is the actual embodiment of what program **C** does – to take a program and use it as its own input).

This interesting construction presents a familiar logical challenge. If program **C**, given its own code as input, exhibits consciousness and halts, then program **C**, given its own code as input, does not exhibit consciousness. If program **C**, given its own code as input, does not exhibit consciousness, then program **C**, given its own code as input, exhibits consciousness. Both cases appear to be logical contradictions. One way out of this contradiction is to assume that the program block that exhibits consciousness does not exist. Another way out is to consider that the supposition that program **c** exists is false; that it is impossible to detect consciousness. This could happen if sometimes program **c** does not give an answer.[f]

---

[f] A third way out of this contradiction would be an inconsistent theory of consciousness encoded into program **c**; either a theory which states a not-conscious program is conscious, or a theory that states a conscious program is not conscious. A fourth way out is if program **c** itself exhibits consciousness.

Program **c** could be implemented as a British Museum Algorithm which generates well-formed proofs in an orderly manner until it finds a proof that its input program **P** operating on input **I** is either conscious or is not conscious. This, of course, would require a theory of consciousness involving the interaction of material things in a manner that can be described mathematically. If such a theory is ever discovered, it could be used to generate program **c**. Note that if program **c** is itself conscious, then program **C**, given its own code as input, could correctly and without contradiction answer that it has consciousness, even though it is fleeting consciousness. Unless such a theory of consciousness concludes that a straightforward implementation of a British Museum Algorithm is itself conscious, program **c** is indeed non-conscious. However, a British Museum Algorithm theory of consciousness is not a compelling theory.

Given a consistent theory of machine consciousness, either machine consciousness is impossible or it is undetectable by a non-conscious machine.

One could argue that this is not much of an argument against machine consciousness.  Similar arguments prove that a computer program cannot inspect another program to predict if it will halt or if it will print some specific output, yet we do know that some programs halt and that some programs print specific things.  So perhaps some programs do exhibit consciousness.  More insight can be found be taking a step back and looking at the larger implications in a way similar to the step back taken in considering Gödel's famous Incompleteness Theorem.

At a higher, or meta-level, program **C** on input **C** cannot determine if its self is conscious.  Yet by definition a conscious entity knows that it is conscious.  Therefore program **C** on input **C** is not conscious.  Thus there is at least one non-conscious program that cannot be identified as non-conscious by a theory of consciousness involving the interaction of material things in a manner that can be described mathematically.   And if a given program cannot be proven to be non-conscious, it follows that not all conscious programs can be proven to be conscious.

Consciousness is undecidable by any formal theory of material consciousness.  Any theory of material consciousness would be incomplete.

**Implications**

Materialists believe as a matter of faith:

- Anything that can be observed is explainable.
- Any explanation of observed phenomena only involves material things.
- All interactions of material things can be quantified mathematically.

Therefore:

- Consciousness is explainable.
- Consciousness only involves material things.
- The interactions resulting in consciousness can be quantified mathematically.

However, consciousness is undecidable by any material theory of consciousness. Gödel proved that mathematical truths exist that cannot be proven.  This extension to Gödel's insight proves that if indeed conscious programs can exist, not all of them can be proven to be conscious.  Any material theory of consciousness must be either inconsistent or incomplete.

# Postlude

 "Real knowledge is to know the extent of one's ignorance." – The Laudably Declarable Lord Ni

It is tautological that we don't know what we don't know.  We do know, however that there are things that we don't know and that there are things that we cannot ever know.  This is a logical consequence that is based only on the assumption that mathematics is consistent.  If mathematics is not consistent, then everything that we know is untrue.  In either event, we are limited to understand no more than a small fraction of all that is true.

# Denouement

"In faith there is enough light for those who want to believe and enough shadows to blind those who don't." – Blaise Pascal

"Faith is a knowledge within the heart, beyond the reach of proof." – Khalil Gibran

Belief in the Divine has always been considered an act of faith.  Scientific reductionism and materialism are considered by many to be intellectually based and the only sensible worldview; any worldview requiring faith is considered by many to be unscientific.  Yet it takes tremendous leap after leap of faith to adhere to this worldview.  It takes faith to believe that physics can explain all natural phenomena.  It takes faith to believe that every phenome could be produced by mutation and natural selection.  It takes faith to believe that consciousness is a manifestation of material interactions.

Some people use the word "faith" like it is a bad thing.  The only time faith is a bad thing is when it is used to deny the decidable.   But when something cannot be proven, and when it is proven that this something is unprovable, then it is appropriate to employ faith.

Some choose to have faith in man.

Some choose to have faith in God.

"You believe what you want to believe."

# List of Figures